\documentclass[runningheads]{sig-alternate}

\usepackage{amssymb}
\usepackage{graphicx}
\usepackage{verbatim}
\usepackage[latin1]{inputenc}
\usepackage{paralist}
\usepackage{subfigure}
\begin{document}

\conferenceinfo{NDA'08,} {October 31, 2008, Fairfax, Virginia, USA.} 
\CopyrightYear{2008}
\crdata{978-1-60558-301-3/08/10} 

\title{The Risk-Utility Tradeoff for IP Address Truncation}

\numberofauthors{2}
\author{
  \alignauthor Martin Burkhart, Daniela Brauckhoff, Martin May \\
  \affaddr{Computer Engineering and Networks Laboratory}   \\
  \affaddr{ETH Zurich, Switzerland }
  \email{\{burkhart,brauckhoff,may\}@tik.ee.ethz.ch}
  \alignauthor Elisa Boschi \\
  \affaddr{Hitachi Europe}   \\
  \email{elisa.boschi@hitachi-eu.com}
}

\maketitle

\begin{abstract}

Network operators are reluctant to share traffic data due to security and privacy concerns.
Consequently, there is a lack of publicly available traces for validating and generalizing
the latest results in network and security research. Anonymization is a possible
solution in this context; however, it is unclear how the sanitization of data preserves
characteristics important for traffic analysis. In addition, the privacy-preserving property of state-of-the-art
IP address anonymization techniques has come into question by recent attacks that successfully identified a large number of hosts in
anonymized traces.

In this paper, we examine the tradeoff between data utility for anomaly detection and the risk of
host identification for IP address truncation.
Specifically, we  analyze three weeks of unsampled and non-anonymized
network traces from a medium-sized backbone network to assess data utility.
The risk of de-anonymizing individual IP addresses is formally evaluated, using a metric
based on conditional entropy.

Our results indicate that truncation effectively prevents host identification but degrades the utility of data for anomaly detection.
However, the degree of degradation depends on the metric used and whether network-internal or external addresses are considered.
Entropy metrics are more resistant to truncation than unique counts and the detection quality of anomalies degrades much faster in
internal addresses than in external addresses. In particular, the usefulness of internal address counts is lost even for truncation of only 4 bits
whereas utility of external address entropy is virtually unchanged even for truncation of 20 bits.

\end{abstract}


\section{Introduction}
\label{sec:intro}


The sharing of network traffic traces is a crucial prerequisite for fostering progress
in network and security research. Unfortunately, even when data export is restricted to packet
headers, as it is the case with Cisco NetFlow, a certain amount of personal information may still be
extracted and exploited to profile user behavior. This
threat to user privacy has already been recognized by data protection legislation in both
Europe \cite{eu-privacy1,eu-privacy2} and the United States~\cite{Ohm:2007:Legal}.
As a result, multiple anonymization tools that aim to prevent the leakage of privacy information have been developed, such as
FLAIM~\cite{Slagell:2006:Flaim}, TCPdpriv~\cite{tcpdpriv}, and CryptoPAn~\cite{Fan:2004:Prefix}. Despite the widespread application of these tools, the effect of the implemented techniques is not yet understood in-depth.

For researchers with access to non-anonymized data sets, this is
not an issue. Unfortunately, only few research institutes have such traffic traces available
and the large majority works with publicly available, but already anonymized data sets. For instance, the widely-used traces from Abilene apply
truncation of 11 bits. Hence, studies on the impact of anonymization methods are needed. Anonymization
techniques need to be evaluated along two dimensions: (i) the residual risk involved in publishing data and (ii) the utility of anonymized data for various applications.

As for the study of risk, recent work has shown that many state-of-the-art techniques for IP address anonymization are not as secure as expected~\cite{Coull:2007:Inferring,Koukis:2006:Risks,Brekne:2005:Attacks,Ribeiro:2008:Prefix}.
The reason for this weakness is rooted in the fact that random permutation and (partial)
prefix-preserving permutation~\cite{Fan:2004:Prefix, Pang:2006:tcpmkpub} are reversible. Permutations
are vulnerable to fingerprinting attacks and behavioral analysis, i.e., individual hosts can be profiled and mapped back to original entities.
Truncation of IP addresses, on the other hand, involves a significant amount of
information loss that thwarts host profiling. We argue that permutation-based anonymization of IP addresses is not sufficient
and propose the use of truncation, which offers a stronger level of privacy by aggregating individual hosts. We formally evaluate
the risk of host identification in truncated flow traces and show that truncation provides stronger privacy guarantees than other anonymization techniques such as permutations.

The remaining question is how truncation preserves data utility for different applications. Being an important application of flow traces, we evaluate trace utility with regard to network anomaly detection, in this paper represented by a Kalman filter approach. The specific problem we are investigating has not yet been addressed in the literature.

Our contributions are the following:
(i) we quantify the utility of truncated data for backbone anomaly
detection with the help of a three-week long data set from a medium-size
ISP and an anomaly detector based on a Kalman filter;
(ii) we derive a metric for the risk of host de-anonymization when
truncation is applied; and
(iii) based on these results, we present a quantitatively evaluated risk-utility map for truncation.

In Section~\ref{sec:related} we briefly discuss related work for risk and utility assessment of anonymization techniques.
Section~\ref{sec:methodology} covers the applied methodology. We then argue that the effect of truncation is different for internal and external addresses, due to an observed asymmetry in prefix structure (Section~\ref{sec:asymmetry}). Utility and risk for truncation are quantified in Sections~\ref{sec:utility} and~\ref{sec:risk}. The findings are consolidated in Section~\ref{sec:rumap}, where a quantitative risk-utility map for truncation is presented.

Among other results, we found that the entropy of addresses is more resistant to truncation than unique address counts. Furthermore, our results show a fundamental asymmetry between internal and external address distributions. With increasing number of truncated bits, both, utility and disclosure risk drop faster for internal than for external addresses.
Finally, we discuss our findings and conclude the paper in section \ref{sec:conclusion}.

\section{Related Work}
\label{sec:related}

A graphical representation of the risk-utility tradeoff for anonymization techniques was introduced by Duncan et al.~\cite{Duncan:2001:RU-Map} with the R-U map, which plots the risk versus utility for anonymization techniques.
We will use the R-U map to summarize our findings from Sections~\ref{sec:utility} and~\ref{sec:risk}.

As stated above, recent attacks have revealed that privacy of individual hosts is in danger when (partial) prefix-preserving permutation is used.
For instance, Ribeiro et al.~\cite{Ribeiro:2008:Prefix} attack prefix-preserving permutation by fingerprinting hosts based on their active ports and exploiting
the structure of the prefix tree. Koukis et al.~\cite{Koukis:2006:Risks} recognize anonymized webservers by means of characteristic object sizes and systematic port-scanning. Brekne et al.~\cite{Brekne:2005:Attacks} analyze the frequency of objects and Coull et al.~\cite{Coull:2007:Inferring} construct behavioral profiles using dominant state analysis. 

All of these attacks have in common the identification of hosts by means of unique characteristics or behavior. These kind of attacks is always feasible when a host-preserving anonymization scheme is used. Hence, even stronger techniques that do not preserve prefixes, such as renumbering of hosts with integer numbers, are vulnerable. In worst case, an attacker capable of injecting traffic into the network prior to anonymization could send individual fingerprint packets to target hosts and later identify them in the anonymized traces. Therefore, we advocate truncation as a remedy against host identification, as it aggregates individual hosts along with their characteristics. This effect of truncation is similar to \textit{k-anonymity}~\cite{sweeney2002kam}. A data set is k-anonymous if individual records are not distinguishable from $k-1$ other records by a set of so-called quasi-identifiers that can be linked to external information. 

While the success of each of these attacks could be used as a metric for assessing host identification risk, some more general metrics have been proposed.
Coull et al.~\cite{Coull:2008:Taming} define a similarity metric for datasets based on information entropy and mutual information. This allows the authors to capture the statistical indistinguishability of objects in the traces and identify high risk objects. However, their analysis is quadratic in the number of objects and has been evaluated only on a small dataset (<30k flows). Bezzi~\cite{Bezzi:2007:Entropy} proposed metrics based on conditional entropy that measure the uncertainty about original records, given the anonymized traces. Based on Bezzi's metrics, ~\cite{Kounine:2008:Assessing} evaluated different anonymization policies, again on a small dataset with roughly 60k flows. Due to the large amount of flows in our dataset (see Section~\ref{sec:data}) and the computational complexity of these metrics, we derive a similar metric based on conditional entropy that formally estimates the host identification risk for truncation (see Section~\ref{sec:risk}). The metric is then parameterized with our data set.

With respect to utility,
Soule et al.~\cite{Soule:2007:Anomalies} study anomaly detection in
NetFlow data from two backbone networks (Abilene and GEANT) that apply different
sampling and anonymization schemes. Although anonymization is not the focus of their
study, they assume that it has an impact on their results. This is exactly where our work concentrates on.
We study the effect of anonymization on raw data that has not been falsified by
prior anonymization or sampling.
The problem of data loss due to anonymization was also identified in~\cite{Zhang:2006:Outsourcing},
where the authors give qualitative recommendations for
anonymization of NetFlow logs when security services are outsourced.
Yurcik et al.~\cite{yurcik2008pat} analyze single-field
anonymization tradeoffs with regard to intrusion detection.
Unfortunately, their dataset contains already anonymized IP
addresses, hence the impact of IP address anonymization techniques
on utility is not studied. On the contrary, \cite{Brauckhoff2005}
and \cite{Mai2006} studied the utility of sampled traffic traces
without addressing the effect of anonymization.

We assess the utility of anonymized data by evaluating the sensitivity of a specific detector.
Our detector models normal traffic with a Kalman filter (as proposed by \cite{Soule2005}), and applies a threshold to the
residual difference between the measured and the predicted signal. We expect, however, similar results for other
popular detectors (e.g., \cite{SubspaceMethod04, Wagner:2005:Entropy}) that apply the same metrics for
detection.

\section{Methodology}
\label{sec:methodology}

In this section, we describe our methodology for studying the
risk and utility of flow data anonymized with truncation.
We introduce the data set used in this study, and
describe the methodology for quantifying the risk and
utility for different truncation parameters.

\subsection{Measurement Data}
\label{sec:data}


The data used in this study was captured from the four border routers of the
Swiss Academic and Research Network (SWITCH, AS 559)~\cite{Switch},
a medium-sized backbone operator, connecting several universities
and research labs (e.g., IBM, CERN) to the Internet. The SWITCH IP
address range contains about 2.2 million IP addresses and the
traffic volume varies between 60 and 140 million NetFlow records per
hour under normal conditions.
We analyzed a three-week period (from August 19th to September 10th
2007). This data set contains a variety of anomalies with
diverse characteristics. In total, 43.2 billion flows covering a
volume of 713 Terabytes of traffic were analyzed.
In contrast to previous work, this study is based on un-sampled and
non-anonymized flow data. Such datasets are difficult to obtain but necessary when bias and distortion
in the results are to be minimized.

\subsection{Risk Quantification}

For a sound quantification of privacy risk, one must specify the goal and the capabilities of the attacker as well as a metric that is suitable to comprise success of the attack.

In our study, the goal of the attacker is host identification. That means, for a given anonymized IP address the attacker tries to recover the original address. We assume that the attacker knows the set of IP addresses in the original trace. That is, the attacker tries to map the IP addresses from the anonymized traces to the list of known original addresses. Such a list could, at least partially, be compiled by scanning of active hosts and using public information about well-known sites. These assumptions are not quite as strong but similar to those of Coull et al.~\cite{Coull:2008:Taming} in the sense that they represent a worst case scenario. In their model, the attacker has exact information about the objects (e.g. hosts, servers or users) and knows the distributions in original data. Similar to Coull et al., the focus of our analysis lies on the statistical ability to distinguish objects rather than the practical details of performing an attack.

For estimation of the de-anonymization risk we apply a metric based on \emph{conditional entropy} along with~\cite{Bezzi:2007:Entropy, Kounine:2008:Assessing}. We formally compute the average probability for the attacker to correctly guess the original address from an anonymized address.

\subsection{Utility Quantification}

\begin{figure}[t]
    \centering
  \includegraphics[scale=0.6]{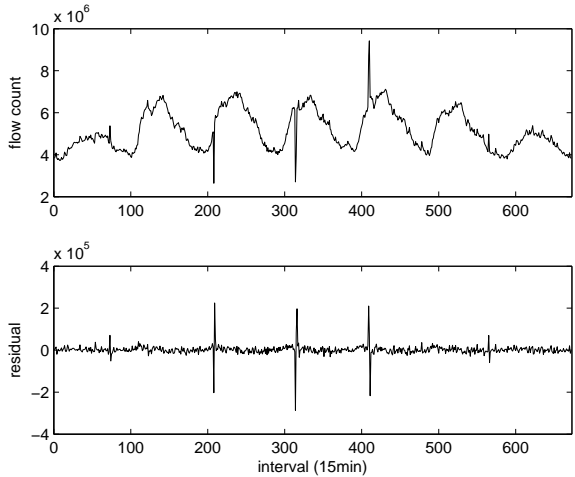}
  \caption{Time series and corresponding residual signal from the Kalman filter.}
  \label{fig:kalman}
  \vspace{-3mm}
\end{figure}

In this Section we describe the methods applied to assess data utility. In particular we describe the studied anonymization technique (truncation), the metrics and procedures used for anomaly detection, and the classification of events in our data.

Utility is very sensitive to the application that anonymized data is used for. Hence, utility always needs to be quantified with a specific application
scenario in mind. The application scenario we are interested in here is the detection of denial of service attacks and network scans in flow data.
Anomaly detection literature has proposed a variety of metrics, e.g., flow, byte, port, address counts, or Shannon entropy, that can be used for the detection of network anomalies (e.g. ~\cite{Wagner:2005:Entropy}). Inspection of our data shows that address counts and entropy are affected by (D)DoS attacks and scans in the following way: 

\begin{description}
\item {\bf(D)DoS}: Denial of Service attacks cause a concentration of the flows on one or few target IP addresses and hence a
drop in the destination IP address entropy. If the attack is distributed or source addresses are spoofed, we additionally see a spike in the
source IP address counts and entropy. 
\item {\bf Scan}: Scans provoke an increase in the destination IP address counts and entropy. If scan
sources are distributed, we also see an increase in the source IP counts and entropy.
\end{description}

Truncation of $x$ bits (e.g., 8 bits) replaces each IP address in a flow trace with its respective $/(32-x)$ (e.g., /24) address prefix.
Hence, truncation mainly impacts metrics based on IP addresses, such as the unique address count, or the distribution of flows per address as captured by the Shannon entropy. Specifically, the number of unique IP address counts is replaced by the number of unique $/(32-x)$ networks in the trace; the distribution of
flows per IP address is replaced by the distribution of flows per $/(32-x)$ network. The question to be addressed here is how this replacement affects anomaly detection results for the respective metrics.

To study this question, we have manually labeled the three-week long flow trace described above. For establishing the ground truth, we computed and visually scanned a variety of volume and entropy metrics commonly applied for anomaly detection, namely flow, packet, and byte counts as well as source/destination address/port counts and entropy. Accordingly, we classified each 15-minutes interval in the three weeks as either normal or anomalous. In a second step, we identified those of the anomalous intervals that have a clear scan or (D)DoS signature. We used only those intervals classified as normal (= 1305 intervals) or scan/(D)DoS event (= 571 intervals) for the evaluation.

To assess the utility of data anonymized with truncation for anomaly detection, we compare the detection rates for IP-based metrics computed on the original traces with those computed on anonymized traces. Specifically, we compare the results for six different truncation rates: no truncation, truncation of 4, 8, 12, 16, and 20 bits. Hence, the metrics we compare are IP address counts, /28, /24, /20, /16 and /12 network counts, as well as entropy per IP addresses, /28, /24, /20, /16 and /12 networks.
Moreover, we distinguish internal and external addresses since we are interested to see whether and how the asymmetry of internal and external prefixes (described in section \ref{sec:asymmetry}) affects the utility.

As detector we apply a Kalman filter (see e.g. ~\cite{Soule2005}) to each metric. The Kalman filter models normal traffic as a ``measurement-corrected'' AR(1) process plus zero-mean Gaussian noise. The difference between this model and the actual measured time series, the so-called residual, is used for detection (see Fig. \ref{fig:kalman} for an illustration). An alarm is raised by the detector if the residual excesses some threshold. To assess the detector performance for each metric and truncation rate, we plot the false positive vs. true positive rates for different thresholds, known as ROC curves. To obtain a single utility value per metric and truncation rate, we compute the Area Under the ROC Curve (AUC)~\cite{Bradley1997}.

\section{Asymmetry of Internal and External Prefixes}
\label{sec:asymmetry}

The impact of anonymization highly depends on the underlying traffic characteristics. For the SWITCH network, we have observed that the distribution of internal and external hosts is quite different. Pang et al.~\cite{Pang:2006:tcpmkpub} make the same distinction by applying a weaker anonymization policy for external than for internal addresses. In their case, the perceived risk is lower for external addresses. In this Section we highlight the distributional differences in our network and explain why we think they give rise to a significant asymmetry between the internal and external domain.

\begin{figure}[t]
  \centering
  \includegraphics[scale=0.55]{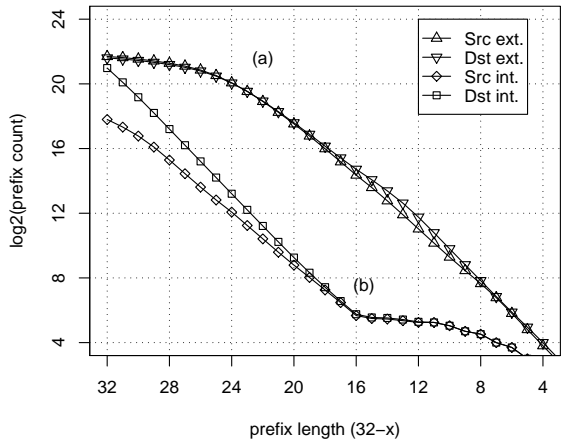}
  \caption{Prefix structure of internal/external hosts for one hour of regular traffic ($x$: truncated bits).}
  \label{fig:prefixStructure}
  \vspace{-4mm}
\end{figure}

As mentioned before, truncation causes a coarse-graining of network entities. Individual hosts are no longer distinguishable because truncation of $x$ bits aggregates all IP addresses in the corresponding /(32-$x$) subnet. As a consequence, metrics based on IP addresses loose a significant amount of detail. Rather than counting IP addresses or computing the entropy of hosts, metrics are computed on prefixes of length 32-x. Figure~\ref{fig:prefixStructure} illustrates this effect with the example of unique address count. As more and more bits are truncated, the number of distinguishable prefixes is reduced. Prefixes are differentiated with respect to direction (source or destination) and address domain (internal or external). We differentiate the direction because this separates targets of scans that are mainly visible in internal destination addresses. The distinction between internal and external is done using the assigned prefix table of the AS. For instance, if a host within the SWITCH network sends a packet to \texttt{google.com}, the host will appear in the internal source addresses and \texttt{google.com} in external destinations. When \texttt{google.com} replies to the request it is added to external sources and the host to internal destinations. Note that a distinction between internal and external address space only makes sense in a stub AS (e.g. SWITCH) or an organization network. For transit ASes (e.g. Abilene), all addresses should be considered external.

Surprisingly, two groups of prefixes with different behavior can be clearly identified in Fig.~\ref{fig:prefixStructure}: those that represent external addresses and those of SWITCH-internal addresses. For increasing $x$, the number of external prefixes remains roughly constant up to turning point \textbf{(a)} around prefix length /24, and decreases exponentially for $x > 8$ bits. In contrast, the number of internal prefixes falls exponentially with $x$ and levels off at turning point \textbf{(b)} around 16 bits.
The plateau before turning point \textbf{(a)} indicates that only few hosts in each external /24 subnet exchange traffic with the observed network. Since a corresponding plateau is missing for internal prefixes, we conclude that we see traffic from almost all hosts of internal /24 subnets. Naturally, the total number of internal /24 subnets is limited by the prefix table of SWITCH. The plateau after turning point \textbf{(b)} is due to the fact that the prefix table is dominated by 30 to 40 complete /16 subnets. Thus, the number of internal prefixes for $x>16$ cannot grow exponentially with prefix length as for external prefixes.

This observation leads to the question whether the truncation-induced reduction of risk and utility is asymmetric with respect to internal and external addresses. That is, we hypothesize that anomalies detectable in external address distributions are more resistant to truncation than those of internal distributions because the characteristics of external /32 prefixes are less impaired by truncation of the first 8 bits (see Fig.~\ref{fig:prefixStructure}). On the risk side, internal addresses aggregate faster with increasing $x$ than external addresses, which makes it harder to identify individual internal hosts.

Note that the difference between internal source and destination prefixes is due to scanning activities that activate virtually all internal addresses (about 2 million) in the destination set. That is, even nonexistent hosts appear in the destination set, simply because a packet was sent to them. Thus, by multiplying the set of potential addresses, scanning could, at first glance, complicate the task of an attacker trying to identify a specific host in the data set. However, we believe it is quite simple to detect scanning activity and focus on the real hosts. Therefore, we only consider internal source addresses, representing the active hosts of the network.

Along with the general assessment of risk and utility in presence of truncation, we try to confirm or refute these asymmetry assumptions in the remainder of this paper.

\section{Utility of Truncated Data}
\label{sec:utility}

\begin{figure}[t]
    \centering
  \includegraphics[scale=0.5]{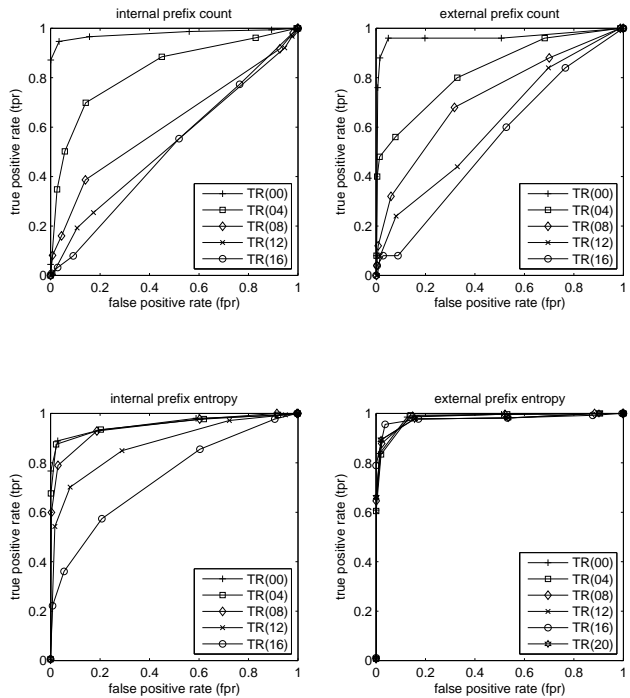}
  \caption{ROC curves for address counts (upper two plots) and address entropy (lower two plots), for internal (left) and external (right) addresses.}
  \label{fig:roc}
  \vspace{-3mm}
\end{figure}

In this section, we present our results regarding the utility of truncated data for the detection of 
scans and denial of service attacks in flow data. In Fig. \ref{fig:roc}, we show the ROC curves for 
two metrics, address counts (upper plots), and address entropy (lower plots), for different truncation rates. Moreover, we distinguish
metrics computed on internal addresses (left-hand side) and external addresses (right-hand side). 
A very accurate detector has a ROC curve that closely follows the left-hand border (low false positive rate) and the top border (high true positive 
rate) of the ROC space; the closer the ROC curve comes to the diagonal of the ROC space the worse becomes the detectors' accuracy.
In Fig. \ref{fig:auc}, we additionally plot the AUC value for each ROC curve for different truncation rates. The area under the curve
(AUC) is a single-value summary of a ROC curve. An AUC value between 0.9 and 1 means excellent detection capabilities, 
while an AUC value of 0.5 means the detector is useless, i.e. not better than random guessing.

\subsection{Counts vs. Entropy}

\begin{figure}[t]
    \centering
  \includegraphics[scale=0.55]{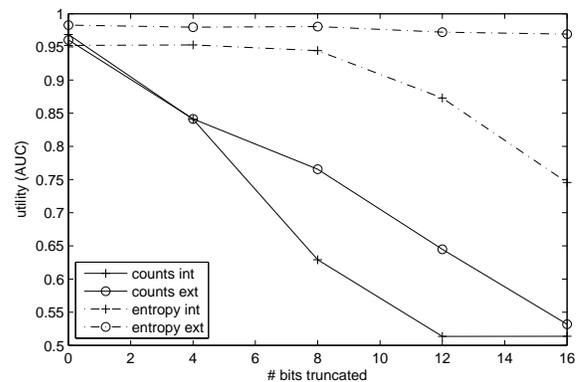}
  \caption{Utility in terms of AUC for the four ROC curves in Fig.~\ref{fig:roc}.}
  \label{fig:auc}
  \vspace{-4mm}  
\end{figure}

When comparing the detection results for count and entropy metrics (upper vs. lower plots in Fig.~\ref{fig:roc} and
dash-dotted vs. solid lines in Fig. \ref{fig:auc}) in truncated data, we see 
a clear difference between the two. While the detection rate on non-anonymized data is high, likewise for 
counts and entropy, the picture changes completely when truncation is used. The detector performance for 
counts decreases significantly with higher truncation rates. Even for a truncation rate as low as 4 bits, 
the detection (or true positive) rate for a given false positive rate of 0.05 is reduced by more than 50\%, which makes detection practically impossible. 
For entropy metrics, on the other hand, the performance decreases much slower. Both, internal and external address entropy, perform well even when the data is anonymized with truncation of as much as 8 bits.

How can we explain this remarkable difference? We have two possible explanations for the bad performance 
of counts: First, it is possible that the attacks are not well distributed with respect to network prefixes, e.g. /16 networks.
A second reason might be that the background noise is higher for prefix counts than for IP address counts. 
Further analysis is required to check on these assumptions. 
The superior performance of entropy metrics on truncated data suggests that the distribution of flows per network prefix is 
similar to the distribution per IP address. The effect of truncation on entropy metrics is that the flows to 
all IP addresses within the same prefix are merged into a single value. Hence, the distribution within the prefix range gets 
lost, but the distribution of flows over all prefix ranges is retained.

We conclude from this observation that the increased complexity for computing entropy metrics, compared to simple 
address counts, is worth the effort, when working with anonymized data. Entropy has already been shown to work reasonably well on the Abilene traces that are anonymized with truncation of 11 bits \cite{SubspaceMethod04, SubspaceMethodSensitivity}.

\subsection{Internal vs. External Prefixes}

We have introduced the prefix asymmetry observed in stub networks, like the SWITCH network we are studying, in section \ref{sec:asymmetry}.
We have shown that almost all available internal prefixes appear in normal traffic (exponential decay of unique prefixes with decreasing prefix length), 
while only a small part of all external prefixes is present in normal traffic (almost the same number of IP addresses and /24 networks).
When comparing the plots for internal and external addresses for count and entropy metrics, we find that count metrics 
(upper row) do not show a significant difference for internal and external addresses, while for entropy metrics (lower row)
there is a clear difference between the two. Entropy computed on external addresses is not affected at all, even when as much as 20 bits
are truncated from the IP addresses.\footnote{The 20 bits rate is only shown for external entropy in Fig.~\ref{fig:roc}.} For internal addresses performance decreases significantly when more than 12 bits are truncated.

We believe that for count metrics the impact of truncation dominates any impact that the difference in 
internal and external addresses could have. For entropy metrics, on the other hand, the distribution is preserved
much better for external addresses with increasing truncation rates. This can be explained with the fact that
the used external address space is much sparser than the internal one, and thus less IP addresses are merged into 
a single value through truncation, and the distribution is less affected.

Therefore, when using entropy as metric, the truncation rate can be up to 20 bits for external addresses without affecting the utility 
of the data for the detection of (D)DoS attacks and network scans, while for internal addresses truncation of 8 bits is acceptable.
These specific results are, however, only valid for our data set, and need to be validated with other sources in the future. In particular, tolerable truncation of internal addresses is highly dependent on the network size.

\section{Risk of Host Identification}
\label{sec:risk}

\begin{table}
	\centering
		\begin{tabular}{|c|c|c|c|}
			\hline
			Original ($S$) & Truncated ($R$) & $P(S|r)$ & $H(S|r)$ \\
			\hline
			129.132.80.15  & 129.132.80.0  & $\frac{1}{3}, \frac{1}{3}, 0, 0, \frac{1}{3},0 $ & 1.6 \\
			129.132.80.77  & 129.132.80.0  & $\frac{1}{3}, \frac{1}{3}, 0, 0, \frac{1}{3},0 $ & 1.6 \\
			129.132.115.5  & 129.132.115.0 & $ 0, 0, \frac{1}{2}, 0, 0, \frac{1}{2}         $ & 1.0 \\
			152.88.3.90    & 152.88.3.0    & $ 0, 0, 0, 1, 0, 0 $                             & 0.0 \\
			129.132.80.144 & 129.132.80.0  & $\frac{1}{3}, \frac{1}{3}, 0, 0, \frac{1}{3},0 $ & 1.6 \\
			129.132.115.90 & 129.132.115.0 & $ 0, 0, \frac{1}{2}, 0, 0, \frac{1}{2} $         & 1.0 \\
			\hline
		\end{tabular}
		\caption{Examples of truncated IP addresses (8 bits) with conditional entropy.}
		\label{tab:risk}
		\vspace{-3mm}
\end{table}

We estimate the risk of host identification by applying a metric based on \emph{conditional entropy} along with~\cite{Bezzi:2007:Entropy, Kounine:2008:Assessing}. Let $R$ be the set of anonymized addresses and $S$ the set of all original addresses. We are now interested in the probability that, given an anonymized address $r \in R$, the correct original address $s \in S$ is recovered. The conditional probability distribution $P(S|r)$ assigns to each $s$ a probability $P(s|r)$ to be the correct original address of an anonymized address $r$. Using the distribution  $P(S|r)$ we now define the conditional entropy on $S$:

\begin{equation}
	H(S|r) := - \sum_{s\in S} P(s|r) \cdot \log_2\big( P(s|r) \big)
	\label{eq:condentropy}
\end{equation}

This expression measures the risk at the level of a a single address $r$. It represents the average number of binary questions that have to be asked to correctly identify the correct $s$ given $r$. Large entropy indicates a bigger uncertainty about the original address whereas low entropy corresponds to an almost deterministic mapping. Table~\ref{tab:risk} gives an example with 8 bits truncation. For each original address $s_i$ in the first column, column 2 holds the anonymized address $r_i$. In the third column,  $P(S|r)$ is shown. Consider for instance, the first anonymized address 129.132.80.0. That is, we now need to calculate $P(s_i|r=129.132.80.0)$ for each $s_i$ in the first column. As there are 3 possible addresses with the prefix 129.132.80.*, the probability for each of these is $1/3$ and $0$ for all addresses with a different prefix. This results in

\begin{displaymath}
	P(S|r=129.132.80.0) = \{\frac{1}{3}, \frac{1}{3}, 0, 0, \frac{1}{3},0\}.	
\end{displaymath}

From this distribution, the conditional entropy is calculated in the last column. Note that with address number 4, the mapping to the original address is deterministic and the corresponding entropy is $0.0$:

\begin{displaymath}
	P(S|r=152.88.3.0) = \{0, 0, 0, 1, 0, 0\}.	
\end{displaymath}

From this local risk measure we derive a global measure: the \emph{average probability of correctly guessing the real identity} of anonymized addresses ($p_{cg})$. Note that there are various ways of defining global risk. Please refer to ~\cite{Bezzi:2007:Entropy} for a discussion. For a specific $r\in R$, $p_{cg}(r)$ equals to $1/{2^{H(S|r)}}$ because $2^{H(S|r)}$ is the number of choices for $r$. Accordingly, the global $p_{cg}$ that averages over all $r$ resolves to

\begin{equation}
	p_{cg}(S,R) := \frac{1}{|R|} \sum_{r\in R} \frac{1}{2^{H(S|r)}}.
	\label{eq:conditional}
\end{equation}

\subsection{Risk for Truncation}

Based on the definition of $p_{cg}$ we now derive a risk metric for truncation of $x$ bits. With truncation of $x$ bits, addresses represent /(32-x) networks. All addresses belonging to the same /(32-x) network are indistinguishable in the truncated trace. If all addresses are active (i.e. present in the traces), this leads to a $P(S|r)$ that has probability $0$ for all $s$ that do not belong to the same /(32-x) network as $r$ and probability $1/{2^x}$ for the $2^x$ values of $s$ that belong to $r$'s network.

However, it is usually not the case that all hosts are active in a network. We assume that our attacker knows $S$ and thus is able to restrict guesses to active hosts. We account for this reduction of host number by introducing a constant $A$ ($A=[0,1]$) that is specific to a network and denotes the fraction of active hosts compared to the assigned address space. The effect of $A$ is that the number of eligible guesses is reduced from $2^x$ to $2^xA$. We end up with a $P(S|r)$ that has $2^xA$ non-zero probabilities each with value $\frac{1}{2^xA}$. Applying this to (\ref{eq:condentropy}) leads to

\begin{eqnarray}
	H_x(S|r) & = & -2^xA\big(\frac{1}{2^xA} \log_2(\frac{1}{2^xA})\big) \\
	         & = & -\big(\log_2(2^{-x}) + \log_2(A^{-1})\big) \\
	         & = & x + \log_2(A). \label{eq:hx}
\end{eqnarray}

This result is consistent with intuition. As we truncate $x$ bits from IP addresses, the number of binary questions to ask for recovering the information is exactly $x$. Restricting the set to a fraction $A$ of possibilities reduces the uncertainty by $-\log_2(A)$. Note that $\log_2(A)$ is always negative as $A<=1$.
Using (\ref{eq:hx}) in (\ref{eq:conditional}), the final risk $p_{cg}$ amounts to

\begin{equation}
  p_{cg}(x) = \left\{
  							\begin{array}{ll}
									\frac{1}{2^xA} & \textrm{if $2^xA>=1$} \\
									1              & \textrm{else}
							\end{array} \right.
	\label{eq:pcg}
\end{equation}

The term $2^xA$, denoting the average number of choices for a given address, could actually be smaller than $1$ due to a scarce network (e.g. no truncation is applied and $A=0.1$). However, we always have at least one potential original address for each anonymized address. Therefore, $p_{cg}$ is set to $1$ for $2^xA<1$. That is, one can not be more than absolutely sure about a mapping.

Note that additional permutation of truncated prefixes in $R$ does not improve worst case privacy of hosts. The goal of metric $p_{cg}$ is to capture statistical distinguishability by means of host characteristics and behavior. As stated in Section~\ref{sec:related}, permutations are reversible using fingerprint attacks. Thus, in the worst case, an attacker could first reverse the permutation but would then again be confronted with the indistinguishability of hosts.

It is important to note that external sources of information could help to further distinguish between hosts. Consider, for example, an attacker that knows the distribution of $S$ in Table~\ref{tab:risk}. That is, he might know that $129.132.115.90$ has ten times more traffic than $129.132.115.5$. Thus, $H(S|129.132.115.0)$ would not be $1.0$ as in the example, but be reduced to $0.44$. The above analysis assumes that no external knowledge is given and that $P(S|r)$ is uniformly distributed among hosts with the same prefix, leading to maximum entropy. Therefore it provides a lower bound estimation of risk. Additional external information affects our analysis by skewing the distribution $P(S|r)$ for blocks of IP addresses, reducing $H(S|r)$ and, in turn, raising the risk $p_{cg}$. 

\subsection{Risk Asymmetry}

As we have already discussed in Section~\ref{sec:asymmetry}, internal and external prefixes exhibit different behavior when truncation is applied (see Fig~\ref{fig:prefixStructure}). Furthermore, the above risk definition gives rise to another noteworthy asymmetry between internal and external address space.

\begin{figure}[t]
    \centering
  \includegraphics[scale=0.55]{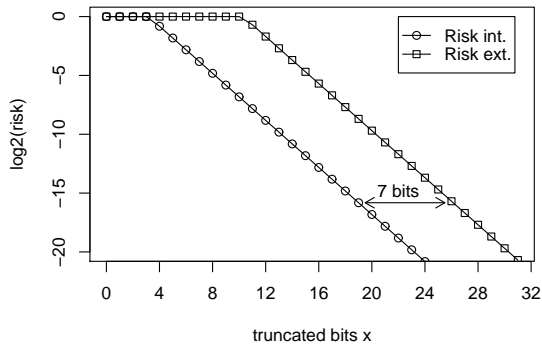}
  \caption{Semi-log plot of internal and external host identification risk $p_{cg}(x)$.}
  \label{fig:risk}
  \vspace{-3mm}
\end{figure}

For SWITCH-internal addresses, $A$ is 10.5\% during the hour shown in Fig.~\ref{fig:prefixStructure}. From the 2.2 million IP addresses assigned to SWITCH, about $230,000$ are active. For external addresses, $A$ is only about 0.08\% (3.4 million visible versus $2^{32} = 4.3\cdot 10^9$ potential addresses). Consequently, external addresses are about 130 times more likely to be correctly de-anonymized than internal addresses, simply because internal addresses are more densely packed and unique identification is harder. To compensate this asymmmetry, one must truncate roughly 7 bits more from external addresses than from internal addresses. Truncation of 7 additional bits leads to a risk reduction by a factor of $2^{7}=128$. Figure~\ref{fig:risk} plots the internal and the external risk versus the truncated bits $x$. Risk falls below $1$ after 3 truncated bits for internal and 10 bits for external addresses. Beyond that point both risks decay exponentially with a constant gap of 7 bits.

Although the 7 bits are specific to the SWITCH network, a similar effect always occurs when traffic is measured at the borders of a network and a distinction between internal and external address space is made. The fraction of active internal hosts tends to be higher because all traffic from internal hosts passes through the borders while only an insignificant portion of the whole internet traffic is seen. In general, the difference in truncated bits for equal risk is $\log_2(A_{in}/A_{out})$.

It is interesting to ask whether it actually makes sense to protect external addresses in the same way as internal addresses. Whereas our results indicate that external addresses should be anonymized stronger, Pang et al.~\cite{Pang:2006:tcpmkpub} explicitly apply the weaker prefix-preserving permutation to external and the stronger partial prefix-preserving permutation to internal addresses. They argue that external addresses are more difficult to attack due to their non-locality. However, with truncation, it is exactly this non-locality that makes external addresses stand out within their networks and makes them uniquely identifiable. 

It is comprehensible that data publishers are more concerned with privacy of their own network and customers. Anyway, data protection laws do not distinguish between internal or external addresses. In particular, in Switzerland and in the European Union any data that can be used either directly or indirectly, i.e., through the use of additional information, to identify an individual is considered ``personal data'' and must therefore be protected. In Switzerland, where our data has been collected, ``personal data'' include data that can be used to identify a legal person, i.e., a company.

In the future, we expect more and more network traces to be published, which is the goal of anonymization research in the first place. Of course, our internal addresses are external addresses in all other data sets. Consequently, if external addresses were anonymized weakly, they could be correlated over different data sets. That would leave people incapable of protecting their own addresses.

\section{The Risk-Utility Tradeoff}
\label{sec:rumap}

In this Section we summarize the results from Sections~\ref{sec:utility} and~\ref{sec:risk} using the R-U map~\cite{Duncan:2001:RU-Map}. The R-U map
is a parametric plot with a running parameter, having x-coordinates (Utility) and y-coordinates (Risk) determined by a function of the parameter. In our case, the parameter is the number of truncated bits $x$, taking the values 0-16 in steps of 4. Depending on $x$, the risk is quantified by Eq.~\ref{eq:pcg} in Section~\ref{sec:risk} and the utility is quantified by the AUC values shown in Fig.~\ref{fig:auc}.

\begin{figure}[t]
  \centering
  \includegraphics[scale=0.55]{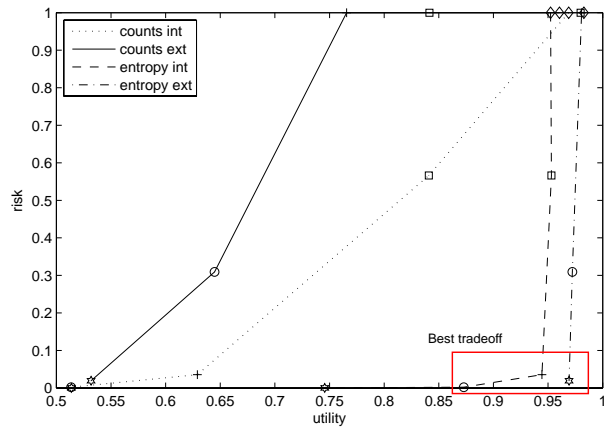}
  \caption{R-U map that illustrates the risk-utility tradeoff for IP address truncation. Markers indicate truncated bits: diamond (no truncation), square (4), plus (8), circle (12), star (16).}
  \label{fig:rumap}
\end{figure}

In Fig. \ref{fig:rumap}, we plot the utility vs. risk for internal and external count and entropy metrics for different 
truncation rates (identified by markers). The utility on the x-axis ranges from 0.5 to 1, where 0.5 corresponds to 
a low detection rate and 1 to a high detection rate. The risk ranges from 1 to 0, where 1 is related to a very high 
risk and 0 to a low risk. Consequently, the sweet spot on this map is the lower right corner with a high
utility and low risk. 

To our surprise, there are three risk-utility combinations within this area of interest, two for internal 
entropy and one for external entropy. Table~\ref{tab:best} summarizes these metrics with the best risk-utility tradeoff.

\begin{table}[t]
	\centering
		\begin{tabular}{|c|c|r|r|}
		  \hline
		  Metric			    	& $x$ &  Utility & Risk \\
		  \hline
			internal entropy  & 8  &  0.94    & 0.035 \\
			internal entropy  & 12  &  0.87    & 0.002 \\
			external entropy  & 16  &  0.97    & 0.020 \\
			\hline
		\end{tabular}
		\caption{Anonymized metrics with the best risk-utility tradeoff ($x$: truncated bits).}
		\label{tab:best}
		 \vspace{-3mm}
\end{table}

For external entropy, truncation of 12 bits also has a very high utility, but the associated risk of more than 0.3 is unacceptable. As expected, both internal and external counts are far off the sweet spot. For counts, as soon as risk drops, utility goes down as well.

\section{Conclusion}
\label{sec:conclusion}

In this paper, we have presented a quantitative assessment of the risk-utility tradeoff involved in IP address truncation.
We evaluated the utility of anonymized data by performing anomaly detection on original and anonymized data. The risk of host identification
in anonymized data was formally estimated using a metric based on conditional entropy. The results were summarized in a comprehensive Risk-Utility map.

We found that entropy metrics are far more resistant to truncation than unique count metrics. Moreover, there is a fundamental asymmetry between
internal and external address distributions for stub networks. Both risk and utility decay faster for internal addresses with increasing number of truncated bits. Consequently, the least useful metric is internal address count as even truncation of only 4 bits renders it useless. On the other hand, the utility of external address entropy is virtually untouched even for truncation of 20 bits. With respect to the risk-utility tradeoff we identified three metrics (calculated on truncated data) with an excellent tradeoff having a remaining utility between 87\% and 97\% and host identification risk of only 0.2\% to 3.5\%. We conclude that truncation offers a better protection against host identification risk than permutation-based IP address anonymization techniques. At the same time, a major part of utility for anomaly detection can be retained using entropy metrics.

\section*{Acknowledgments}
We are grateful to SWITCH for providing the traffic traces used in this study.

\bibliographystyle{abbrv}

\balancecolumns

\end{document}